\documentclass[aps,prl,twocolumn,showpacs,superscriptaddress,nofootinbib]{revtex4}

\usepackage{amsmath}   % need for subequations
\usepackage{graphicx}  % for figures
\usepackage{xspace}
\usepackage{units}

\usepackage{hyperref}

\newcommand{\minerva}{MINERvA\xspace}
\newcommand{\minos}{MINOS\xspace}

\newcommand{\numubar}{\ensuremath{\bar{\nu}_{\mu}}\xspace}
\newcommand{\dsdq}{\ensuremath{d\sigma/dQ^2}\xspace}

\newcommand{\footnoteremember}[2]{\footnote{#2}\newcounter{#1}\setcounter{#1}{\value{footnote}}}
\newcommand{\footnoterecall}[1]{\footnotemark[\value{#1}]}

\newcommand{\sizecheck}{0} % 0 to do nothing; 1 to check size
\newcommand{\PRLsupp}{0}   % 0 to do nothing; 1 to put the appendix in a supplement
\ifnum\PRLsupp=0
  \newcommand{\SuppLocation}{in the Appendix}
\else
  \newcommand{\SuppLocation}{at URL}
\fi

\newif\ifpdf
\ifx\pdfoutput\undefined
   \pdffalse
\else
   \pdfoutput=1
   \pdftrue
\fi
\ifpdf
   \usepackage{graphicx}
   \usepackage{epstopdf}
\else
   \usepackage{graphicx}
\fi

\begin{document}

%\title{Measurement of $d\sigma/dQ^2$ in Muon Antineutrino Quasi-Elastic Scattering on a Hydrocarbon Target with the MINER$\nu$A Detector at Energies Between 1.5 and 10 GeV} 
\title{Measurement of Muon Antineutrino Quasi-Elastic Scattering \\ on a Hydrocarbon Target at $E_{\nu} \sim$ 3.5~GeV} 
%Energies Between 1.5 and 10 GeV} 

%\title{Measurement of \dsdq in Muon Antineutrino Quasi-Elastic Scattering on a %Hydrocarbon Target}

%% MANUAL PARTS OF AUTHOR LIST

%% (1) need to add ``\thanks{\deceased}'' after DeMaat, Gobbi, Tzanakos
\newcommand{\deceased}{Deceased}

%% (2) we offered Jan authorship, so here it is
%% have to put his author line in by hand
\newcommand{\wroclaw}{Institute of Theoretical Physics, Wroc\l aw University, Wroc\l aw, Poland}    
\newcommand{\Rutgers}{Rutgers, The State University of New Jersey, Piscataway, New Jersey 08854, USA}
\newcommand{\Hampton}{Hampton University, Dept. of Physics, Hampton, VA 23668, USA}
\newcommand{\Dortmund}{Institute of Physics, Dortmund University, 44221, Germany }
\newcommand{\Otterbein}{Department of Physics, Otterbein University, 1 South Grove Street, Westerville, OH, 43081 USA}
\newcommand{\JMU}{James Madison University, Harrisonburg, Virginia 22807, USA}
\newcommand{\Florida}{University of Florida, Department of Physics, Gainesville, FL 32611}
\newcommand{\UCIrvine}{Department of Physics and Astronomy, University of California, Irvine, Irvine, California 92697-4575, USA}
\newcommand{\CBPF}{Centro Brasileiro de Pesquisas F\'{i}sicas, Rua Dr. Xavier Sigaud 150, Urca, Rio de Janeiro, RJ, 22290-180, Brazil}
\newcommand{\PUCP}{Secci\'{o}n F\'{i}sica, Departamento de Ciencias, Pontificia Universidad Cat\'{o}lica del Per\'{u}, Apartado 1761, Lima, Per\'{u}}
\newcommand{\INRM}{Institute for Nuclear Research of the Russian Academy of Sciences, 117312 Moscow, Russia}
\newcommand{\Jlab}{Jefferson Lab, 12000 Jefferson Avenue, Newport News, VA 23606, USA}
\newcommand{\Pittsburgh}{Department of Physics and Astronomy, University of Pittsburgh, Pittsburgh, Pennsylvania 15260, USA}
\newcommand{\Guanajuato}{Campus Le\'{o}n y Campus Guanajuato, Universidad de Guanajuato, Lascurain de Retana No. 5, Col. Centro. Guanajuato 36000, Guanajuato M\'{e}xico.}
\newcommand{\Athens}{Department of Physics, University of Athens, GR-15771 Athens, Greece}
\newcommand{\Tufts}{Physics Department, Tufts University, Medford, Massachusetts 02155, USA}
\newcommand{\WM}{Department of Physics, College of William \& Mary, Williamsburg, Virginia 23187, USA}
\newcommand{\FNAL}{Fermi National Accelerator Laboratory, Batavia, Illinois 60510, USA}
\newcommand{\Purdue}{Department of Chemistry and Physics, Purdue University Calumet, Hammond, Indiana 46323, USA}
\newcommand{\MCLA}{Massachusetts College of Liberal Arts, 375 Church Street, North Adams, MA 01247}
\newcommand{\UMD}{Department of Physics, University of Minnesota -- Duluth, Duluth, Minnesota 55812, USA}
\newcommand{\Northwestern}{Northwestern University, Evanston, Illinois 60208}
\newcommand{\UNI}{Universidad Nacional de Ingenier\'{i}a, Apartado 31139, Lima, Per\'{u}}
\newcommand{\Rochester}{University of Rochester, Rochester, New York 14610 USA}
\newcommand{\Austin}{Department of Physics, University of Texas, 1 University Station, Austin, Texas 78712, USA}
\newcommand{\USM}{Departamento de F\'{i}sica, Universidad T\'{e}cnica Federico Santa Mar\'{i}a, Avda. Espa\~{n}a 1680 Casilla 110-V, Valpara\'{i}so, Chile}
\newcommand{\Geneva}{University of Geneva, Geneva, Switzerland}
\newcommand{\Chicago}{Enrico Fermi Institute, University of Chicago, Chicago, IL 60637 USA}
\newcommand{\keppelThanks}{\thanks{now at the Thomas Jefferson National Accelerator Facility, Newport News, VA 23606 USA}}
\newcommand{\giulianoThanks}{\thanks{now at Vrije Universiteit Brussel, Pleinlaan 2, B-1050 Brussels, Belgium}}
\newcommand{\LazaThanks}{\thanks{also at Department of Physics, University of Antananarivo, Madagascar}}
\newcommand{\schulteThanks}{\thanks{now at Temple University, Philadelphia, Pennsylvania 19122, USA}}
\newcommand{\jwaldingThanks}{\thanks{now at Dept. Physics, Royal Holloway, University of London, UK}}

% 89 total signatories.
\author{L.~Fields}                        \affiliation{\Northwestern}
\author{J.~Chvojka}                       \affiliation{\Rochester}
\author{L.~Aliaga}                        \affiliation{\WM}  \affiliation{\PUCP}
\author{O.~Altinok}                       \affiliation{\Tufts}
\author{B.~Baldin}                        \affiliation{\FNAL}
\author{A.~Baumbaugh}                     \affiliation{\FNAL}
\author{A.~Bodek}                         \affiliation{\Rochester}
\author{D.~Boehnlein}                     \affiliation{\FNAL}
\author{S.~Boyd}                          \affiliation{\Pittsburgh}
\author{R.~Bradford}                      \affiliation{\Rochester}
% ...
%% END AUTOMATIC PART
\collaboration{The \minerva  Collaboration}\ \noaffiliation

\date{\today}

\begin{abstract}
We have isolated \numubar charged-current
quasi-elastic interactions occurring in the segmented
scintillator tracking region of the \minerva detector running in the NuMI 
neutrino beam at Fermilab.  We measure the flux-averaged differential 
cross-section, \dsdq, and compare to several theoretical models of 
quasi-elastic scattering. Good agreement is obtained with a model 
where the nucleon axial mass, $M_A$, is set to $\unit[0.99]{GeV/c^2}$ 
but the nucleon vector form factors are modified to account 
for the observed enhancement, relative to the free nucleon case, of the cross-section for the exchange of transversely polarized photons in electron-nucleus scattering. Our data at higher $Q^2$ favor this interpretation over an alternative in which the axial mass is increased.
\end{abstract}
% Modifying only the axial form-factor by  increasing $M_A$ to $\unit[1.35]{GeV/c^2}$ is in poorer agreement with the data.
\ifnum\sizecheck=0  
\maketitle
\fi

%\section{Introduction}

The recent discovery that the neutrino mixing angle $\theta_{13}\approx 9^\circ$~\cite{Abe:2011sj,Adamson:2013ue,Abe:2011fz,An:2012eh,Ahn:2012nd} makes measuring the hierarchy of neutrino masses and CP violation possible in precision neutrino oscillation experiments.  
Quasi-elastic interactions, $\overline{\nu} p \rightarrow \ell^+  n$ and  $\nu n \rightarrow \ell^- p$, have simple kinematics and serve as reference processes in those experiments ~\cite{AguilarArevalo:2008rc,Abe:2011sj,Ayres:2004js} at GeV energies. These processes are typically modeled as scattering on free nucleons in a relativistic Fermi gas (RFG), with a nucleon axial form factor measured in neutrino-deuterium quasi-elastic scattering~\cite{Bodek:2007vi,Kuzmin:2007kr}. In the RFG model~\cite{Smith:1972xh} the initial state nucleons are independent in the mean field of the nucleus, and therefore the neutrino energy and momentum transfer $Q^2$ can be estimated from the polar angle $\theta_\ell$ and momentum $p_\ell$ of the final state lepton. 
However, correlations and motion of the initial state nucleons, as well as interactions of the final state particles within the nucleus, significantly modify the Fermi gas picture and affect the neutrino energy reconstruction in oscillation experiments~\cite{Martini:2012uc,Lalakulich:2012hs,Nieves:2012yz}. 

Few measurements of antineutrino quasi-elastic scattering exist~\cite{Bonetti:1977cs,Ahrens:1988rr,AguilarArevalo:2013hm}.  
The most recent, from the MiniBooNE experiment on a hydrocarbon target at energies near $1$~GeV~\cite{AguilarArevalo:2013hm}, does not agree with expectations based on the RFG model described above.  A MiniBooNE analysis of $\nu_\mu$ quasi-elastic scattering suggests an increased axial form factor at high $Q^2$~\cite{AguilarArevalo:2007ab}. 
However, results at higher energy from the NOMAD experiment~\cite{Lyubushkin:2008pe} are consistent with the Fermi gas model and the form factor from deuterium.

In this Letter we report the first study of antineutrino quasi-elastic interactions from the \minerva experiment, which uses a finely segmented scintillator detector at Fermilab to measure muon antineutrino and neutrino charged current interactions at energies between $1.5$ and $10$~GeV on nuclear targets.  
The signal reaction has a $\mu^+$ in the final state along with one or more nucleons (typically with a leading neutron), and no mesons\footnote{{I}n this analysis quasi-elastic scattering occurs on both free protons and inside carbon nuclei.}. 
The $\mu^+$ is identified by a minimum ionizing track that traverses  \minerva~\cite{minerva_nim} and travels downstream to the MINOS magnetized spectrometer~\cite{Michael:2008bc} where its momentum and charge are measured. 
The leading neutron, if it interacts, leaves only a fraction of its energy in the detector in the form of scattered low energy protons.  
To isolate quasi-elastic events from those where mesons are produced, we require the hadronic system recoiling against the muon to have a low energy. That energy is measured in two spatial regions. The {\it vertex energy} region corresponds to a sphere around the vertex with a radius sufficient to contain a proton (pion) with 120 (65)~MeV kinetic energy. This region is sensitive to low energy protons which could arise from correlations among nucleons in the initial state or interactions of the outgoing hadrons inside the target nucleus. We do not use the vertex energy in the event selection.  The {\it recoil energy} region includes energy depositions outside of the vertex region and is sensitive to pions and higher energy nucleons. We use the recoil energy to estimate and remove inelastic backgrounds.

%\section{\minerva Experiment and Data} 

The \minerva experiment studies neutrinos produced in the NuMI beamline~\cite{Anderson:1998zza} from \unit[120]{GeV} protons which strike a graphite target.   The mesons produced in $p+C$ interactions are focused by two magnetic horns into a \unit[675]{m} long helium-filled decay pipe.  The horns were set to focus negative mesons, resulting in a muon antineutrino enriched beam with a peak energy of \unit[3]{GeV}.  Muons produced in meson decays are absorbed in \unit[240]{m} of rock downstream of the decay pipe. This analysis uses data taken between November 2010 and February 2011 with $1.014\times 10^{20}$ protons on target.

A Geant4-based~\cite{1610988,Agostinelli2003250} beamline simulation 
is used to predict the antineutrino flux.  Hadron production in the simulation 
was tuned to agree with the NA49 measurements of pion production
from 158~GeV protons on a thin carbon target~\cite{Alt:2006fr}. FLUKA is used to
translate NA49 measurements to proton energies between 
12 and \unit[120]{GeV}~\cite{Ferrari:2005zk,Battistoni:2007zzb}.  
Interactions not constrained by the NA49 data are predicted using the 
FTFP hadron shower model\footnote{{F}TFP shower model in Geant 4 version 92 patch 03.}.

The \minerva detector consists of a core of scintillator
strips surrounded by electromagnetic and hadronic calorimeters on the
sides and downstream end of the detector\footnote{The \minerva scintillator tracking region is 95\% CH and 5\% other materials by weight.}~\cite{minerva_nim}.  The strips are perpendicular to the z-axis (which is very nearly the beam axis) and are arranged in planes with a 1.7~cm strip-to-strip pitch\footnote{The y-axis points along the zenith and the beam is directed downward by \unit[58]{mrad} in the y-z plane.}. Three plane orientations ($0^\circ, \pm 60^\circ$ rotations around the z-axis) enable reconstruction of the neutrino
interaction point, the tracks of outgoing charged particles, and calorimetric
reconstruction of other particles in the interaction.  The \unit[3.0]{ns}
timing resolution is adequate for separating multiple interactions within a
single beam spill.  

\minerva is located \unit[2]{m} upstream of the MINOS near detector, 
a magnetized iron spectrometer~\cite{Michael:2008bc}.  
The \minerva detector's response is simulated by a tuned 
Geant4-based~\cite{Agostinelli2003250,1610988} program.  
The energy scale of the detector is set by ensuring
that both the photostatistics and the reconstructed energy deposited by
momentum-analyzed through-going muons agree in data and simulation.
Calorimetric constants used to reconstruct the energy of hadronic showers 
are determined from the simulation. The uncertainty in the response 
to single hadrons is constrained by the measurements made with a 
scaled down version of the \minerva detector in a 
low energy hadron test beam~\cite{minerva_nim}.

The \minerva\ detector records the energy and time of energy depositions (hits) in each scintillator strip.  Hits are first grouped in time and then clusters of energy are formed by spatially grouping the hits in each scintillator plane.  Clusters with energy $>\unit[1]{MeV}$ are then matched among the three views to create a track. The most upstream cluster on the muon track establishes the event vertex. We identify a $\mu^+$ by matching a track that exits the back of \minerva with a positively charged track entering the front of \minos. The per plane track resolution is 2.7~mm and the angular resolution of the
muon track is better than 10~mrad~\cite{minerva_nim}.  
 The event vertex is restricted to be within the central 110 planes of the scintillator tracking region and no closer than \unit[22]{cm} to any edge of the planes. These requirements define a region with a mass of \unit[5.57]{metric tons}.

%The event vertex is restricted to a hexagon of apothem 85~cm centered  on the \minerva detector and must be within the central 110 planes  of the fully active detector.  These requirements define a region  with a mass of 5.57~tons.

The times of the tracked hits are used to determine the interaction time.  
Other untracked clusters up to \unit[20]{ns} before and 
\unit[35]{ns} after that time are associated with the event.  The energy of the 
recoil system  is calculated from all clusters not associated with the muon
track or located within the vertex region.
Events with two or more isolated groups of
spatially contiguous clusters 
%of recoil energy 
are rejected as likely to be due to 
inelastic backgrounds.

Event pile-up causes a decrease in the muon track reconstruction efficiency. We studied this in both \minerva and \minos by projecting tracks found in one of the detectors to the other and measuring the misreconstruction rate. This resulted in a -7.8\% (-4.6\%) correction to the simulated efficiency for muons below (above) 3 GeV/c.

Estimation of the initial neutrino energy ($E_\nu$) and 
four-momentum transfer squared ($Q^2$) of the
interaction assumes an initial state nucleon at rest with a constant
binding energy, $E_b$, which we set to $\unit[+30]{MeV}$ based on
electron scattering data~\cite{Moniz:1971mt,VanOrden:1980tg} and estimates of Coulomb and asymmetry (Pauli) energy effects from the semi-empirical 
mass formula for nuclei~\cite{Katori:2008zz}. 
Under this quasi-elastic hypothesis,
denoted by $QE$, 
\begin{eqnarray}
E_\nu^{QE} &=& \frac{m^2_n - (m_p-E_b)^2-m_\mu^2+2(m_p-E_b)E_\mu}{2(m_p-E_b-E_\mu+p_\mu \cos\theta_\mu)} \label{eq:enudef} \\
%{\rm\textstyle and}~
Q^2_{QE} &=& 2E_\nu^{QE}(E_\mu-p_\mu \cos\theta_\mu ) - m_\mu^2, \label{eq:q2def}
\end{eqnarray}
where $E_\mu$ and $p_\mu$ are the muon energy and momentum, $\theta_\mu$ is the muon angle
with respect to the beam and $m_n$, $m_p$ and $m_\mu$ are the masses
of the neutron, proton and muon, respectively.

\begin{figure}[tp]
\centering
\ifnum\PRLsupp=0
  \includegraphics[width=\columnwidth]{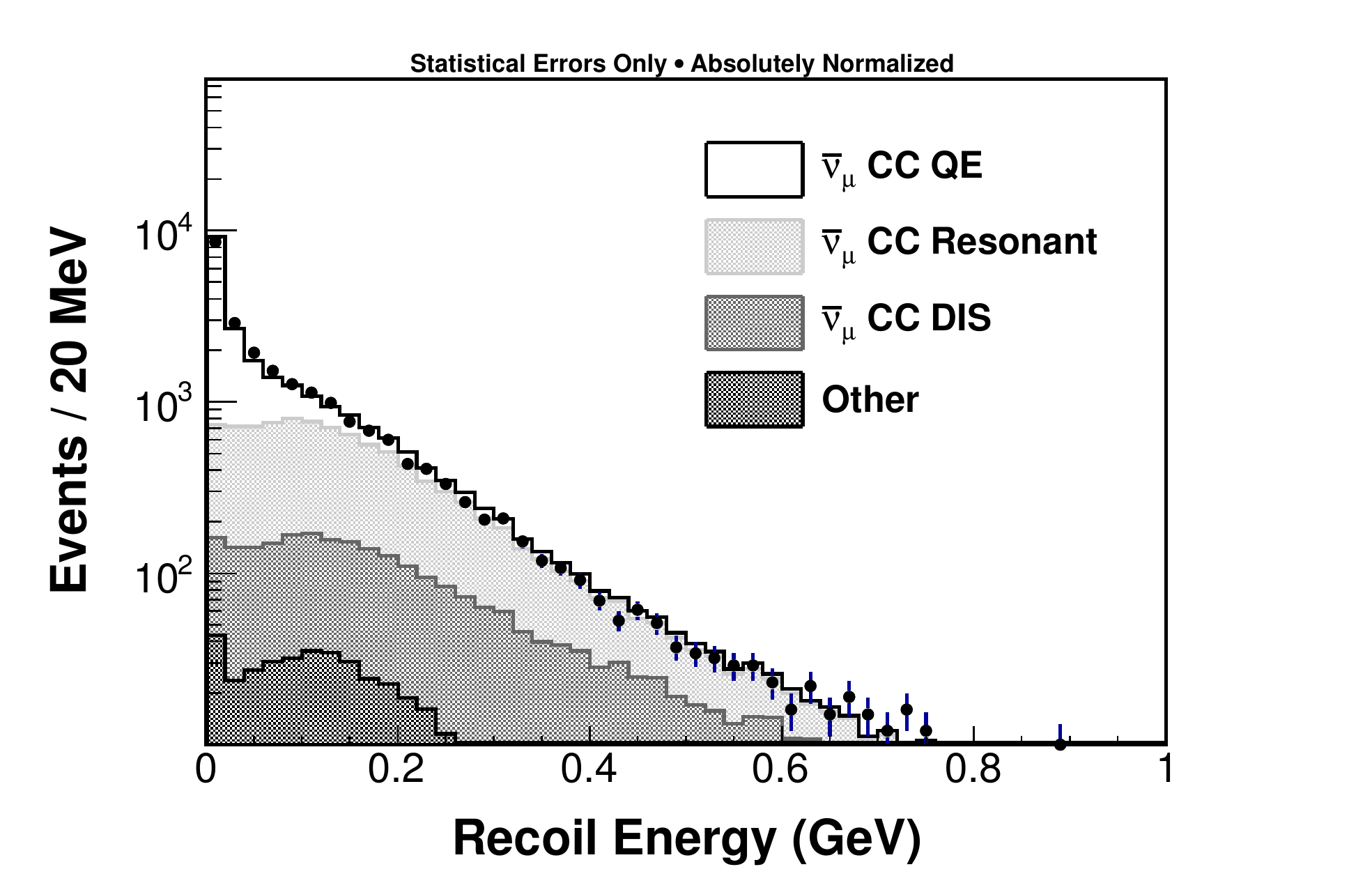} 
\else
  \includegraphics[width=0.8\columnwidth]{figures/Figure1_anu} 
\fi
\caption{The measured recoil energy distribution (solid circles) and the predicted composition of signal and background. Backgrounds from baryon resonance production (light grey), continuum/deep-inelastic scattering (dark grey), and other sources (black),such as coherent pion production, are shown.  The fraction of signal in this sample, before requiring low recoil energy, is $0.58$.}
\label{fig:recoil_fits}
\end{figure}

Figure~\ref{fig:recoil_fits} shows the reconstructed data compared 
to neutrino interactions simulated 
using the GENIE 2.6.2 neutrino event generator~\cite{Andreopoulos201087}.  
For quasi-elastic interactions,  the cross-section is given by the 
Llewellyn Smith formalism~\cite{LlewellynSmith:1971zm}.  
Vector form factors come from fits 
to electron scattering data~\cite{Bradford:2006yz}; 
the axial form factor used is a dipole with an axial mass 
($M_A$) of 0.99~GeV$/$c$^2$, consistent with 
deuterium measurements~\cite{Bodek:2007vi,Kuzmin:2007kr}; and
sub-leading form factors are assumed from PCAC or exact G-parity
symmetry~\cite{Day:2012gb}.  The nuclear model is the relativistic
Fermi gas (RFG) 
with a Fermi momentum of $221$~MeV$/$c and an extension to higher 
nucleon momenta to account for short-range 
correlations~\cite{Bodek:1980ar,Bodek:1981wr}.
Inelastic reactions with a low invariant mass hadronic final state are based on a tuned model of discrete
baryon resonance production~\cite{Rein:1980wg}, and the transition to deep inelastic scattering is simulated using the Bodek-Yang
model~\cite{Bodek:2004pc}.   Final state interactions, where hadrons
interact within the target nucleus, are modeled using the INTRANUKE
package~\cite{Andreopoulos201087}.

\begin{figure}[tp]
\centering
\ifnum\PRLsupp=0
  \includegraphics[width=\columnwidth]{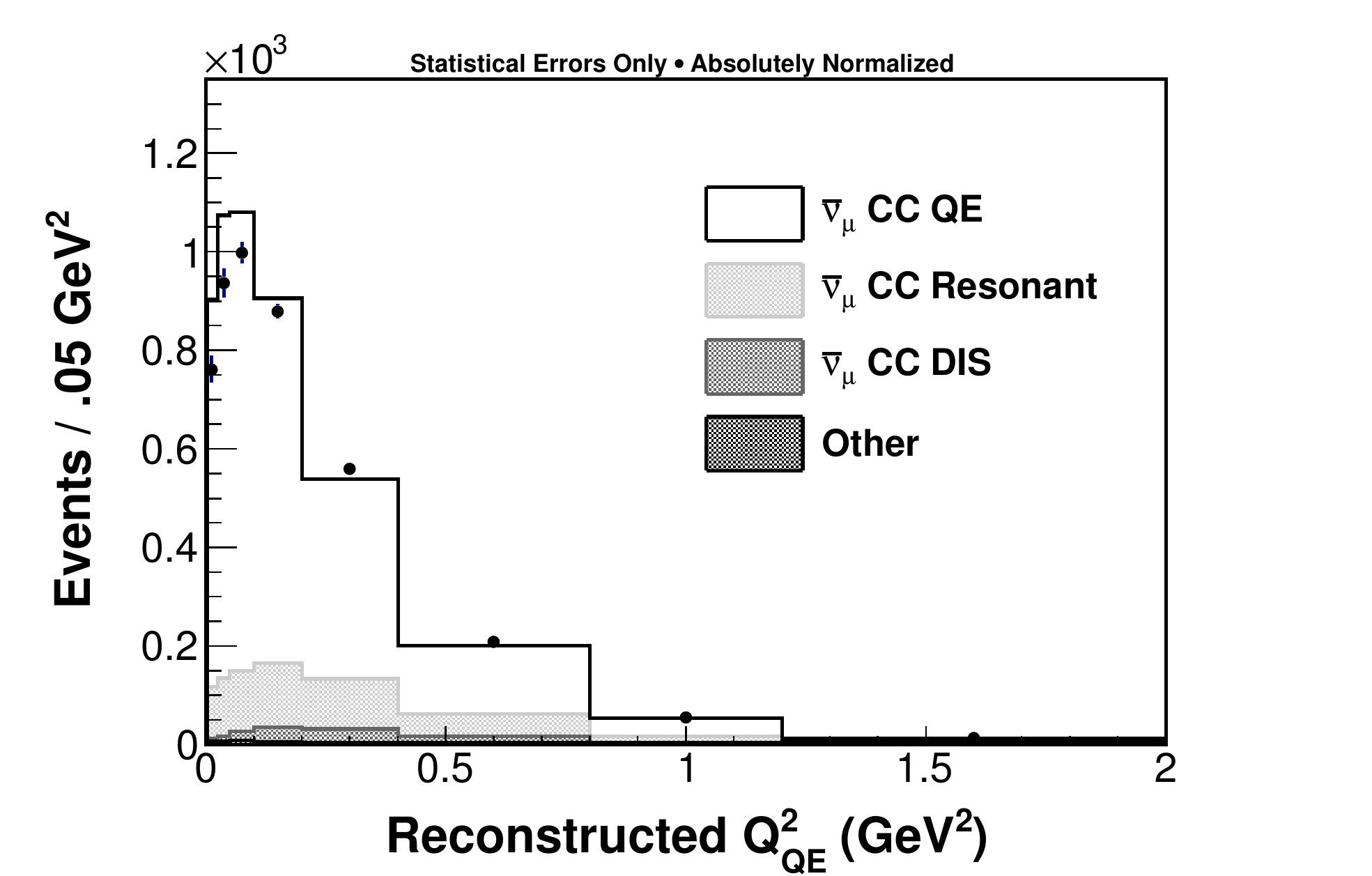} 
\else
  \includegraphics[width=0.7\columnwidth]{figures/Figure2_anu} 
\fi
\caption{\label{fig:qsq} The measured $Q^2_{QE}$ distribution before background subtraction and corrections for detector resolutions and acceptance.  The fraction of signal in this sample is $0.77$, and 54\% of signal events in our fiducial volume pass all selections.}
\end{figure}

Figure~\ref{fig:recoil_fits} shows evidence of quasi-elastic interactions in the
peak of events at low recoil energy. A significant background of 
inelastic events still exists, primarily from baryon resonance 
production and decay where the final state pion is not identified.  To reduce this 
background we make a $Q^2_{QE}$ dependent selection of low recoil-energy events\footnote{{T}he precise selection is $E_\text{recoil} < 0.03+0.3\times Q^2_{QE}$(GeV$^2/$c$^2$). The $Q^2_{QE}$ dependence improves the signal efficiency for higher $Q^2_{QE}$.}.  We also require $E_\nu^{QE} < \unit[10]{GeV}$ to limit uncertainties due to the neutrino flux. Figure~\ref{fig:qsq} shows the $Q^2_{QE}$ 
distribution of the remaining 16,467 events in the data compared with the simulation. 

% MAK, the 1\% background sentence doesn't fit very well here and I'm not sure we need it
% The $\mu^-$ background predicted by the simulation is less than 1\%.

%At this point the signal efficiency is estimated at xx\% and the
%purity of the sample is estimated at yy\% .

%\section{Cross-Section Extraction Procedure} 

%for $Q_{QE}^2>0.75$~GeV$^2$.
The background in each $Q^2_{QE}$ bin is estimated from 
the data by fitting the relative normalizations of signal and background 
recoil energy distributions whose shapes are taken from the simulation. 
The fit results in a 10\% reduction in the relative background estimate for 
 $Q^2_{QE}> \unit[0.8]{GeV^2}$ and no change to $Q^2_{QE} < \unit[0.8]{GeV^2}$.
We then correct for energy resolution using a Bayesian 
unfolding method~\cite{D'Agostini:1994zf} with four iterations to 
produce the event yield as a function of $Q^2_{QE}$, determined via 
Eq.~\ref{eq:q2def}  with $p_\mu$ and $\theta_\mu$ taken 
from the GENIE event generator.  
After unfolding, we use the simulation to correct the yield for efficiency and acceptance, and then divide by the neutrino flux and the number of target nucleons to calculate the bin-averaged cross-section.  We estimate the neutrino flux in the range $1.5\le E_\nu \le \unit[10.0]{GeV}$ to be $\unit[2.43 \times 10^{-8}]{cm^{-2}}$ per proton on target\footnoteremember{footsupp}{See Supplemental Material\ \SuppLocation\ for the flux as a function of energy and for correlations of uncertainties among bins for the cross-section and shape measurement}, and there are  $1.91 \pm 0.03 \times 10^{30}$ protons in the fiducial volume. 

% the error on the number of protons is 1.4%

%After
%unfolding, we correct the number of events as a function of true
%$Q^2_{QE}$ for efficiency estimated in the simulation.
%% There are two effects that may not be correctly modeled in the
%% simulation: one is the tracking and matching efficiency in \minerva
%% and the other is the tracking efficiency in MINOS.  These effects were
%% studied using inclusive antineutrino charged current events in the
%% data and simulation.  In antineutrino running only the latter effect
%% effect is underestimated and a correction of 1.5\% is applied.
%The cross-section is this efficiency corrected $Q^2_{QE}$
%from true muon kinematics divided by the neutrino flux
%and by the number of protons in the fiducial volume of the detector.  
%We estimate the flux integrated between 1.5 and 10~GeV in antineutrino energy %to be $2.429\times 10^{-8}/$cm$^2$ per proton on target\footnote{{A}dd flux in supplemental material to be posted online to accompany paper... note this is also a possibility for the cross-section tables.},  
%and there are
%$1.907\times 10^{30}$ protons in the fiducial volume.

%\section{Systematic Uncertainties} 

\begingroup
\squeezetable
\begin{table}
\begin{tabular}{cccccccc}
$Q^2_{QE}$ (GeV$^2$) & I & II & III & IV & V & VI & Total \\
\hline
$0.0 - 0.025$ & 0.05 & 0.04 & 0.00 & 0.02 & 0.11 & 0.02 & 0.13 \\
$0.025 - 0.05$ & 0.05 & 0.04 & 0.01 & 0.01 & 0.11 & 0.02 & 0.13 \\
$0.05 - 0.1$ & 0.05 & 0.04 & 0.01 & 0.01 & 0.11 & 0.01 & 0.13 \\
$0.1 - 0.2$ & 0.04 & 0.04 & 0.01 & 0.01 & 0.11 & 0.01 & 0.12 \\
$0.2 - 0.4$ & 0.03 & 0.06 & 0.01 & 0.02 & 0.11 & 0.01 & 0.13 \\
$0.4 - 0.8$ & 0.05 & 0.07 & 0.02 & 0.03 & 0.11 & 0.01 & 0.15 \\
$0.8 - 1.2$ & 0.11 & 0.11 & 0.02 & 0.02 & 0.11 & 0.02 & 0.20 \\
$1.2 - 2.0$ & 0.13 & 0.15 & 0.04 & 0.04 & 0.12 & 0.02 & 0.23 \\
\hline
\end{tabular}
\caption{Fractional systematic uncertainties on $d\sigma/dQ^2_{QE}$ associated with muon reconstruction (I), recoil reconstruction (II), neutrino interaction models (III), final state interactions (IV), flux (V) and other sources (VI).  The final column shows the total fractional systematic uncertainty due to all sources.}
\label{tab:systematics}
\end{table}
\endgroup

The main sources of systematic uncertainty in the differential
cross-section measurement are due to: the reconstruction of the muon;  
the reconstruction and detector response for hadrons; 
 the neutrino interaction model; final state interactions; 
and the neutrino flux.  These uncertainties are evaluated by 
repeating the cross-section analysis with systematic shifts 
applied to the simulation and their effect is shown in 
Tab.~\ref{tab:systematics}.

%The second largest uncertainty comes from the muon momentum scale in MINOS. 
Uncertainties in the muon energy scale have a direct impact on $Q^{2}_{QE}$ and result in a bin migration of events. The bulk of the uncertainty comes from \minos, which reconstructs muon energy by range (for stopping tracks) and curvature (exiting). There is a 2.0\% uncertainty in the range measurement, due to the material assay and imperfect knowledge of muon energy loss\cite{Michael:2008bc}. Using muon tracks which stop in the detector, we compare the momentum measured by range and curvature to establish an additional uncertainty of 0.6\% (2.6\%) on curvature measurements above (below) \unit[1]{GeV/c}.
We also account for subdominant uncertainties on the  
 energy loss in \minerva, systematic offsets in the 
beam angle, mis-modeling of the angular and position resolution, 
and tracking efficiencies.

The systematic error on the recoil energy measurement is due to the
uncertainty in the \minerva detector energy scale set by muons and
differences between the simulated calorimetric response 
to single hadrons and the response measured by the test beam program.
Additional uncertainties are due to differences between the Geant model of
neutron interactions and thin target data on neutron scattering in
carbon, iron and copper\cite{Abfalterer:2001gw,Schimmerling:1973bb,Voss:1956,Slypen:1995fm,Franz:1989cf,Tippawan:2008xk,Bevilacqua:2013rfq,Zanelli:1981zz}.   We evaluate further sources of systematic error 
by loosening analysis cuts on energy near the vertex 
and on extra isolated energy depositions, repeating the 
fit to the background and subsequent analysis, and assigning
 an uncertainty to cover the difference.  

Predictions for $Q^{2}_{QE}$ and recoil energy distributions for neutrino-induced background processes are based upon the GENIE generator. We evaluate the systematic error by varying the underlying model tuning parameters according to their uncertainties~\cite{Andreopoulos201087}. These include parameters governing inelastic interactions of neutrinos with nucleons and those that vary the final state interactions.  

The systematic error on the antineutrino flux arises from uncertainties in hadron production in the NuMI target and beamline, and from imperfect modeling of the beamline focusing and geometry~\cite{Pavlovic:2008zz}.  Where hadron production is constrained by NA49 data\cite{Alt:2006fr}, the NA49 measurement uncertainties dominate.  The uncertainty on other interactions is evaluated from the spread between different Geant4 hadron production models~
%\cite{Agostinelli2003250}. 
\cite{Agostinelli2003250,1610988}
The absolute flux uncertainties are large, with a significant $E_\nu$ dependence, but mostly cancel in a measurement of the shape of $d\sigma /dQ^2_{QE}$.

\begin{figure}[tp]
\centering
\ifnum\PRLsupp=0
  \includegraphics[width=\columnwidth]{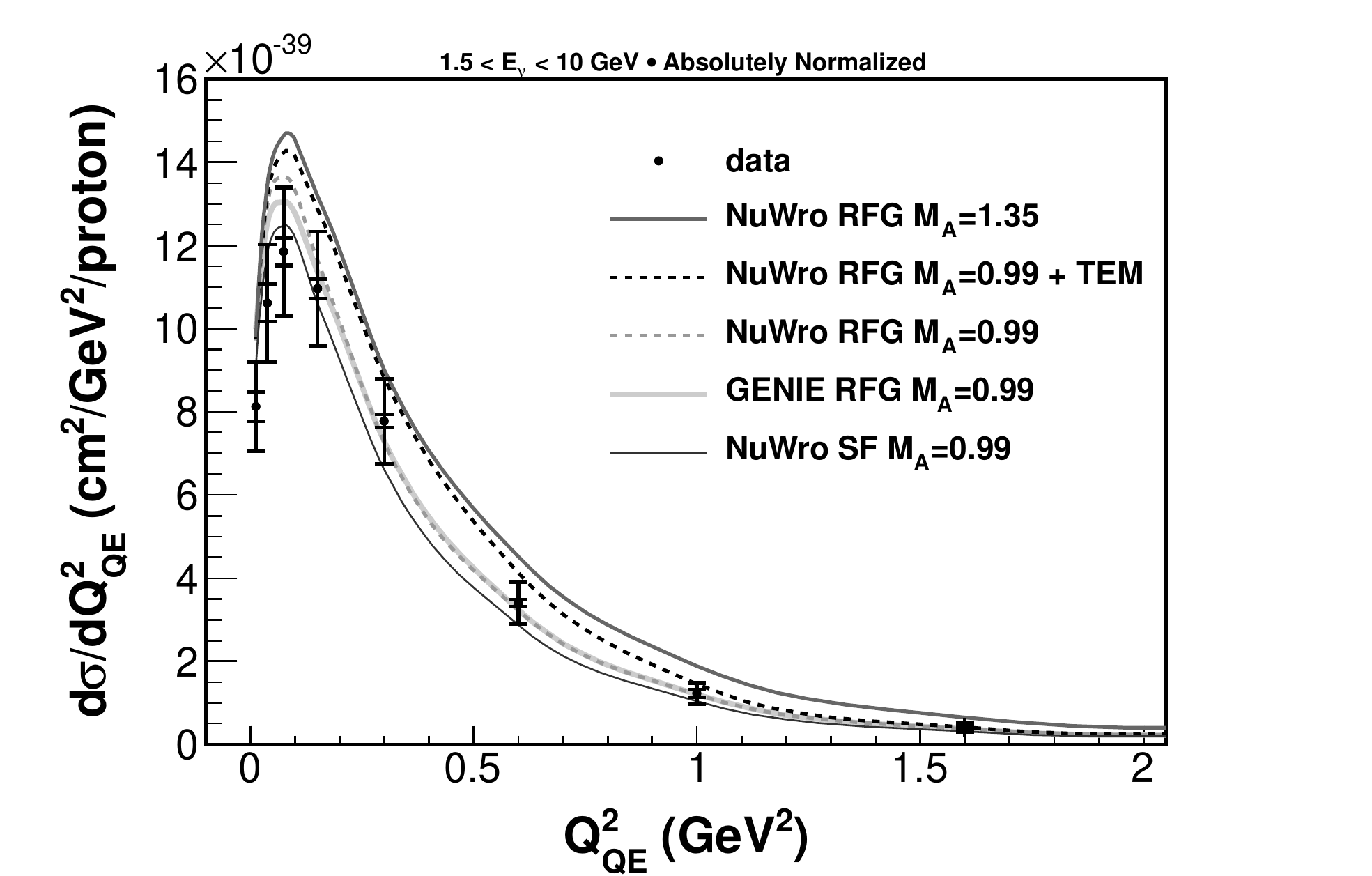} 
\else
  \includegraphics[width=0.8\columnwidth]{figures/Figure3_anu_norpa} 
\fi
%    \vspace{-7pt}
\caption{The anti-neutrino
  quasi-elastic cross-section as a function of $Q^2_{QE}$ compared
  with several different models of the interaction described in the
  text. The inner (outer) error bars correspond to the
  statistical (total) uncertainties.}
%    \vspace{-10pt}
\label{fig:xsec_q2}
\end{figure}

\begin{figure}[tp]
\centering
\ifnum\PRLsupp=0
  \includegraphics[width=\columnwidth]{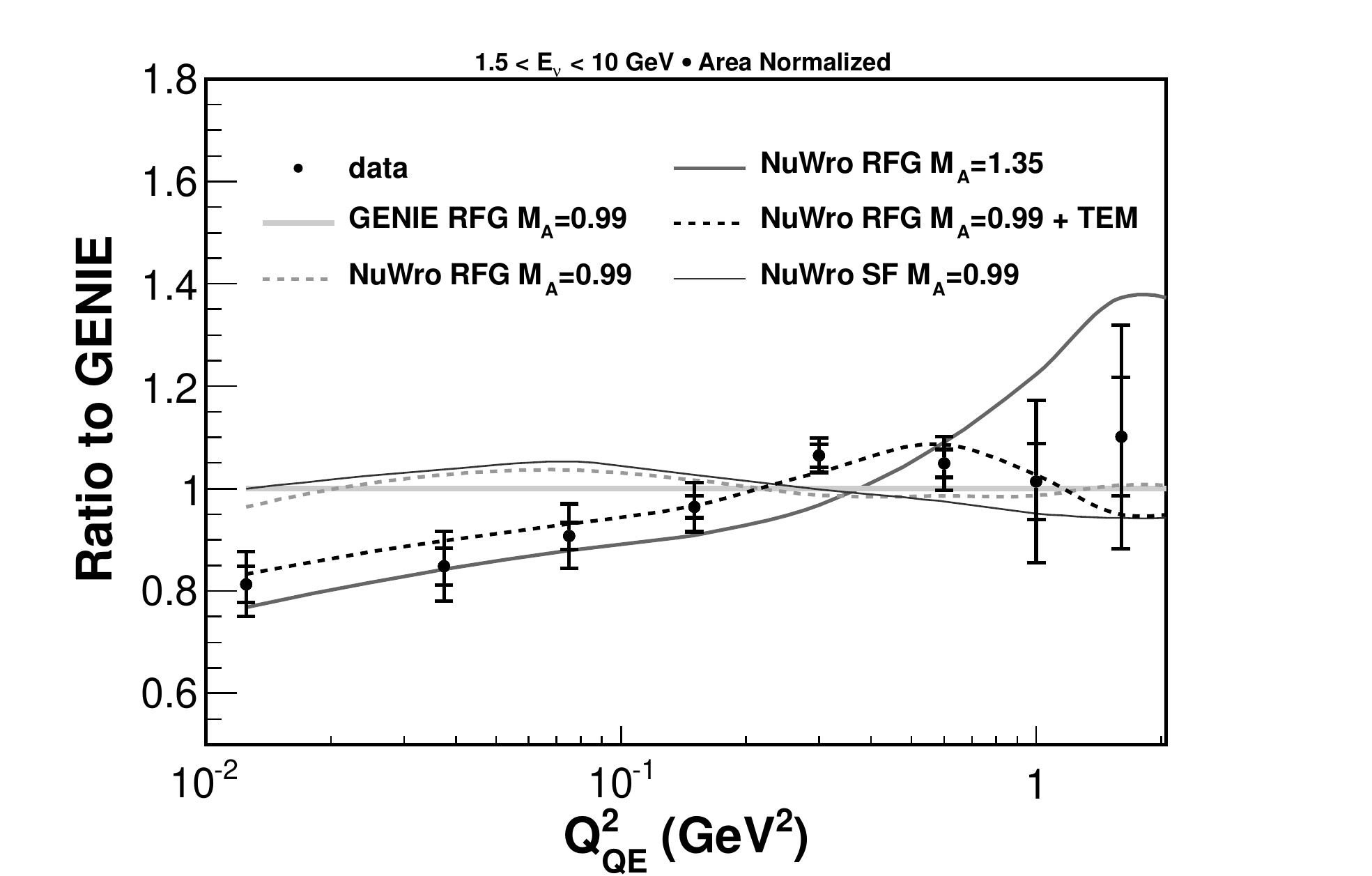} 
\else
  \includegraphics[width=0.8\columnwidth]{figures/Figure4_anu_norpa} 
\fi
\caption{The data and models of Fig.~\ref{fig:xsec_q2} shown by $Q^2_{QE}$ shape and as a ratio to the reference GENIE prediction.}
\label{fig:xsec_q2_shape_ratio}
\end{figure}

The measured differential cross-section $d\sigma/dQ^2_{QE}$ is shown in 
Fig.~\ref{fig:xsec_q2} and Table~\ref{tab:xsec}. 
Averaged over the flux from 1.5 to 10~GeV, we find  $\sigma=0.604 \pm 0.008 \mbox{(stat)} \pm 0.075 \mbox{(syst)} \times \unit[10^{-38}]{cm^2/proton}$.  
As noted above, the systematic uncertainties
are significantly reduced in the shape of the differential
cross-section\footnoterecall{footsupp}, which is shown in Fig.~\ref{fig:xsec_q2_shape_ratio}.
%Figures~\ref{fig:xsec_q2} and \ref{fig:xsec_q2_shape_ratio} 

Table~\ref{tab:chi2} compares the data to the RFG model in the 
GENIE event generator and a number of different nuclear models 
and values of $M_A$ in the NuWro generator~\cite{Golan:2012wx}.
%Models incorporating a low $Q^2_{QE}$ suppression 
%in the random phase approximation (RPA)~\cite{Singh1992587,Graczyk:2003ru} 
%underpredict the cross-section at low $Q^2_{QE}$, and 
There
is little sensitivity to replacement of the Fermi gas with
a spectral function (SF) model of the target nucleon energy-momentum
relationship~\cite{Benhar:1994hw}. The data disfavor $M_A=\unit[1.35]{GeV/c^2}$  as extracted from fits of the MiniBooNE neutrino quasi-elastic data
in the RFG model~\cite{AguilarArevalo:2007ab}. Our data are consistent with a transverse enhancement model (TEM) which has $M_A=\unit[0.99]{GeV/c^2}$ in agreement with deuterium data and includes an enhancement of the magnetic form factors of bound nucleons that has been observed in electron-carbon scattering~\cite{Bodek:2011ps}. The $M_A=\unit[1.35]{GeV/c^2}$ and TEM models have a similar $Q^{2}_{QE}$ dependence at low $Q^{2}_{QE}$ but are distinguished by the kinematic reach of the data at $Q^{2}_{QE}>\unit[1]{GeV^2}$.


\begingroup
\squeezetable
\begin{table}
\begin{tabular}{ccc}
$Q^2_{QE}$ & Cross-section  &  Fraction of \\
{\footnotesize (GeV$^2$) }    & {\footnotesize ($10^{-38}\mathrm{cm}^2/\mathrm{GeV}^2/$proton) } &  Cross-section {\footnotesize (\%)} \\ \hline
$0.0 - 0.025$ & $0.813\pm0.035\pm0.102$ &  $3.45\pm0.15\pm0.22$ \\ 
$0.025 - 0.05$ & $1.061\pm0.045\pm0.134$ &  $4.50\pm0.19\pm0.31$ \\ 
$0.05 - 0.1$ & $1.185\pm0.033\pm0.150$ &  $10.05\pm0.28\pm0.63$ \\ 
$0.1 - 0.2$ & $1.096\pm0.024\pm0.135$ &  $18.59\pm0.41\pm0.83$ \\ 
$0.2 - 0.4$ & $0.777\pm0.016\pm0.101$ &  $26.38\pm0.55\pm0.62$ \\ 
$0.4 - 0.8$ & $0.340\pm0.009\pm0.050$ &  $23.11\pm0.61\pm0.98$ \\ 
$0.8 - 1.2$ & $0.123\pm0.009\pm0.024$ &  $8.35\pm0.61\pm1.15$ \\ 
$1.2 - 2.0$ & $0.041\pm0.004\pm0.010$ &  $5.57\pm0.59\pm0.94$ \\ 
\hline
\end{tabular}
\caption{Table of absolute and shape-only cross-section results.  In each measurement, the first error is statistical and the second is systematic.}
\label{tab:xsec}
\end{table}
\endgroup

\begingroup 
\squeezetable 
\begin{table} 
\begin{tabular}{c|cccc}
NuWro  &  ~RFG~ & ~RFG~  & ~RFG~ & ~SF~ \\  
Model   &          & +TEM~  & &  \\ \hline
$M_A$ (GeV) &  0.99 & 0.99 & 1.35 & 0.99  \\  \hline \hline
%Rate $\chi^2$  & 21.1 & 8.46 & 23.2 & 17.1  \\ 
%Shape $\chi^2$ & 20.3 & 4.59 & 12.1 & 20.9 \\
Rate $\chi^2$/d.o.f. & 2.64 & 1.06 & 2.90 & 2.14 \\
Shape $\chi^2$/d.o.f. & 2.90 & 0.66 & 1.73 & 2.99  \\
%FGM & 0.99 & 33.8 & 30.2 & 33.4 & 36.5 \\ 
%FGM + TEM & 0.99 & 10.3 & 21.7 & 6.29 & 18.1 \\ 
%FGM & 1.35 & 29.3 & 30.7 & 16.9 & 23.6 \\ 
%SF & 0.99 & 28.8 & 24.2 & 34.3 & 34.2 \\ 
%RPA & 0.99 & 24.5 & 75.8 & 27.2 & 91.9 \\ 
\end{tabular}
\caption{\label{tab:chi2} Comparisons between the measured $d\sigma/dQ^2_{QE}$ (or its shape in $Q^2_{QE}$) and different models implemented using the NuWro neutrino event generator, expressed as $\chi^2$ per degree of freedom (d.o.f.) for eight (seven) degrees of freedom.  The $\chi^2$ computation in the table accounts for significant 
correlations between the data points caused by systematic uncertainties. }
%\caption{\label{tab:chi2} $\chi^2$ statistic for eight (seven) degrees of %freedom for comparison between the measured $d\sigma/dQ^2_{QE}$, as well as its %shape in $Q^2_{QE}$, and different models implemented using the NuWro neutrino %event generator. The $\chi^2$ computation accounts for significant 
%correlations between the data points caused by systematic uncertainties.
%}%, for both neutrino and antineutrino modes.
\end{table}
\endgroup

%Transverse enhancement is included as a parametrization affecting the $Q^{2}_{QE}$  dependence in our analysis, however it is thought to be due to underlying multinucleon dynamical processes [49- 54].   Such processes are not included in
%The experiment's simulation;  if present they could effect our measurements of vertex energy and of recoil energy in ways which are not accounted for. 

\begin{figure}[tp]
\centering
\ifnum\PRLsupp=0
  \includegraphics[width=\columnwidth]{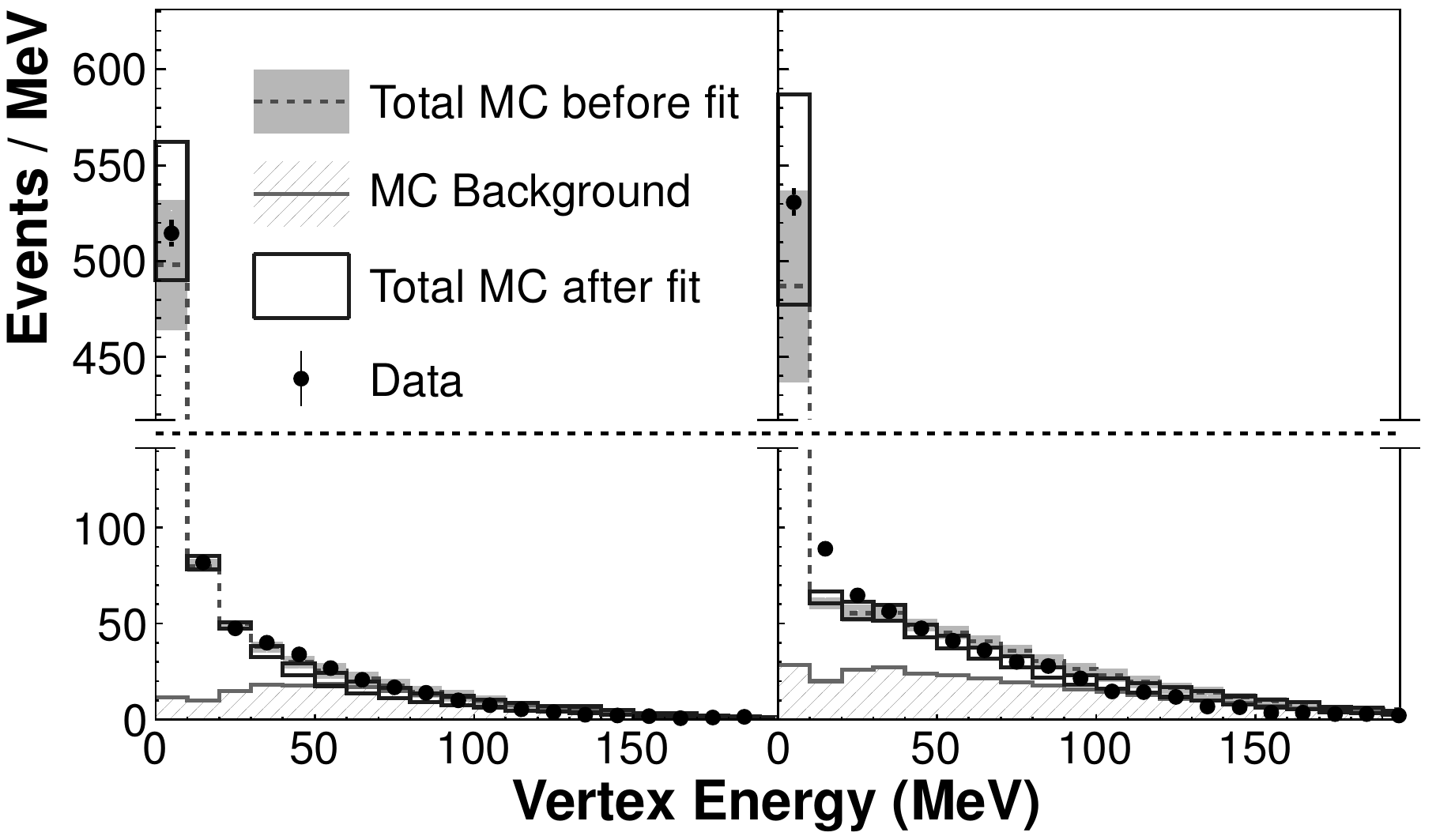}
\else
  \includegraphics[width=0.7\columnwidth]{figures/nubar_vtxE_sharedaxis_horiz}
\fi
\caption{Reconstructed vertex energy of events passing the selection criteria compared to the GENIE RFG model for $Q^2_{QE} < 0.2$ GeV$^2/$c$^2$ (left) and for $Q^2_{QE} > 0.2$ GeV$^2/$c$^2$ (right).}
    \vspace{-10pt}
\label{fig:vtx_eng}
\end{figure}

Transverse enhancement is included as a parametrization affecting the $Q^{2}_{QE}$ dependence in our analysis but is thought to be due to underlying
 multinucleon dynamical processes~\cite{Donnelly:1999sw,Carlson:2001mp,Maieron:2009an,Martini:2009uj,Amaro:2010sd,Martini:2010ex,Nieves:2013fr}. Such processes could have an 
effect on the vertex and recoil energy distributions that we do not simulate. 
Motivated by these concerns and by discrepancies observed 
in our analysis of $\nu_\mu$ quasi-elastic scattering~\cite{nuprl}, 
we have also studied the vertex energy to test the simulation 
of the number of low energy charged particles emitted in 
quasi-elastic interactions. Figure~\ref{fig:vtx_eng} shows this 
energy compared to the simulation.   A fit which modifies the
distributions to incorporate energy due to additional protons is not able 
to achieve better agreement. This might be explained if the dominant multibody 
process is $\bar{\nu}_\mu (n p) \to \mu^{+} n n$~\cite{Donnelly:1999sw,Martini:2009uj,Subedi:2008zz} 
since \minerva is not very sensitive to low energy neutrons. 
A similar analysis on neutrino mode data is consistent with 
additional protons in the final state~\cite{nuprl}.
%and is helpful 
%in drawing further conclusions about the effect of the nucleus on 
%quasi-elastic reactions

%We have done a similar analysis on neutrino mode data which is helpful  in drawing further conclusions about the effect of the nucleus on  quasi-elastic reactions~\cite{nuprl}.

\ifnum\sizecheck=0
  \begin{acknowledgments}

This work was supported by the Fermi National Accelerator Laboratory
%, which is operated by the Fermi Research Alliance, LLC, 
under US Department of Energy contract
No. DE-AC02-07CH11359 which included the \minerva construction project.
Construction support also
was granted by the United States National Science Foundation under
Award PHY-0619727 and by the University of Rochester. Support for
participating scientists was provided by NSF and DOE (USA) by CAPES
and CNPq (Brazil), by CoNaCyT (Mexico), by CONICYT (Chile), by
CONCYTEC, DGI-PUCP and IDI/IGI-UNI (Peru), by Latin American Center for
Physics (CLAF) and by RAS and the Russian Ministry of Education and Science (Russia).  We
thank the MINOS Collaboration for use of its
near detector data. Finally, we thank the staff of
Fermilab for support of the beamline and the detector.

\end{acknowledgments}

  \bibliographystyle{apsrev4-1}
\bibliography{CCQE-nubar}

\fi

\ifnum\PRLsupp=0
  \clearpage
  \newcommand{\qsq}{\ensuremath{Q^2_{QE}}\xspace}
\renewcommand{\textfraction}{0.05}
\renewcommand{\topfraction}{0.95}
\renewcommand{\bottomfraction}{0.95}
\renewcommand{\floatpagefraction}{0.95}
\renewcommand{\dblfloatpagefraction}{0.95}
\renewcommand{\dbltopfraction}{0.95}
\setcounter{totalnumber}{5}
\setcounter{bottomnumber}{3}
\setcounter{topnumber}{3}
\setcounter{dbltopnumber}{3}

\begingroup
\squeezetable
\begin{table*}[!h]
{\normalsize \appendix{Appendix: Supplementary Material}\hfill\vspace*{4ex}}
\tabcolsep=0.11cm
\begin{tabular}{c|cccccccc}
\hline
$\qsq$ (GeV$^2$) Bins & $0.0 - 0.025$ & $0.025 - 0.05$ & $0.05 - 0.1$ & $0.1 - 0.2$ & $0.2 - 0.4$ & $0.4 - 0.8$ & $0.8 - 1.2$ & $1.2 - 2.0$ \\ 
\hline
Cross-section in bin  
 & 0.813 & 1.061 & 1.185 & 1.096 & 0.777 & 0.340 & 0.123 & 0.041 \\ 
($10^{-38}\mathrm{cm}^2/\mathrm{GeV}^2/$proton)  & $\pm$ 0.108 & $\pm$ 0.142 & $\pm$ 0.154 & $\pm$ 0.137 & $\pm$ 0.103 & $\pm$ 0.051 & $\pm$ 0.026 & $\pm$ 0.011\\
\hline
%$\qsq$ (GeV$^2$) & & & & & & & & \\
$0.0 - 0.025$  & 1.000 & 0.884 & 0.911 & 0.901 & 0.828 & 0.700 & 0.362 & 0.297 \\ 
$0.025 - 0.05$  &  & 1.000 & 0.913 & 0.904 & 0.820 & 0.675 & 0.343 & 0.278 \\ 
$0.05 - 0.1$  &  &  & 1.000 & 0.942 & 0.875 & 0.726 & 0.353 & 0.319 \\ 
$0.1 - 0.2$  &  &  &  & 1.000 & 0.933 & 0.825 & 0.431 & 0.413 \\ 
$0.2 - 0.4$  &  &  &  &  & 1.000 & 0.916 & 0.541 & 0.566 \\ 
$0.4 - 0.8$  &  &  &  &  &  & 1.000 & 0.643 & 0.653 \\ 
$0.8 - 1.2$  &  &  &  &  &  &  & 1.000 & 0.752 \\ 
$1.2 - 2.0$  &  &  &  &  &  &  &  & 1.000 \\ 
\hline
\end{tabular}
\caption{The measurement of the differential cross-sections in $Q^2_{QE}$,
their total (statistical and systematic) uncertainties, and the correlation matrix for these uncertainties}
\end{table*}
\endgroup

\begingroup
\squeezetable
\begin{table*}[!h]
\tabcolsep=0.11cm
\begin{tabular}{c|cccccccc}
\hline
$\qsq$ (GeV$^2$) Bins & $0.0 - 0.025$ & $0.025 - 0.05$ & $0.05 - 0.1$ & $0.1 - 0.2$ & $0.2 - 0.4$ & $0.4 - 0.8$ & $0.8 - 1.2$ & $1.2 - 2.0$ \\ 
\hline
\% of cross-section   
& 3.45	& 4.50	& 10.05	& 18.59	& 26.38	& 23.11	& 8.35	& 5.57	\\
in bin
& $\pm	0.27	$ & $\pm	0.36	$ & $\pm	0.69	$ & $\pm	0.93	$ & $\pm	0.83	$  &$\pm	1.15	$ & $\pm	1.30	$ &	$\pm	1.11$ \\
\hline
%$\qsq$ (GeV$^2$) & & & & & & & & \\
$0.0 - 0.025$  & 1.000 & 0.675 & 0.722 & 0.672 & 0.221 & -0.464 & -0.440 & -0.577 \\ 
$0.025 - 0.05$  &  & 1.000 & 0.742 & 0.716 & 0.241 & -0.517 & -0.450 & -0.585 \\ 
$0.05 - 0.1$  &  &  & 1.000 & 0.779 & 0.347 & -0.543 & -0.562 & -0.635 \\ 
$0.1 - 0.2$  &  &  &  & 1.000 & 0.386 & -0.434 & -0.627 & -0.671 \\ 
$0.2 - 0.4$  &  &  &  &  & 1.000 & -0.051 & -0.571 & -0.375 \\ 
$0.4 - 0.8$  &  &  &  &  &  & 1.000 & 0.080 & 0.186 \\ 
$0.8 - 1.2$  &  &  &  &  &  &  & 1.000 & 0.568 \\ 
$1.2 - 2.0$  &  &  &  &  &  &  &  & 1.000 \\ 
\hline
\end{tabular}
\caption{The measurement of the {\em shape} of the differential cross-sections for $Q^2_{QE}<2.0$~GeV$^2$, their total (statistical and systematic) uncertainties, and the correlation matrix for these uncertainties}
\end{table*}
\endgroup

\begingroup
\squeezetable
\begin{table*}[!h]
\tabcolsep=0.11cm
\begin{tabular}{l|ccccccccccccc}
\hline 
$E_\nu$ in Bin & 
$1.5 - 2$ &
$2 - 2.5$ &
$2.5 - 3$ &
$3 - 3.5$ &
$3.5 - 4$ &
$4 - 4.5$ &
$4.5 - 5$ &
$5 - 5.5$ \\
$\bar\nu_\mu$ Flux
(neutrinos/cm$^2$/POT ($\times 10^{-8}$) &
$0.281$	& 
$0.368$	&
$0.444$	&
$0.448$	&
$0.349$	&
$0.205$	&
$0.106$	&
$0.061$	\\
\hline
$E_\nu$ in Bin & 
$5.5 - 6$ &
$6 - 6.5$ &
$6.5 - 7$ &
$7 - 7.5$ &
$7.5 - 8$ &
$8 - 8.5$ &
$8.5 - 9$ &
$9 - 9.5$ &
$9.5 - 10$ \\
$\bar\nu_\mu$ Flux
(neutrinos/cm$^2$/POT ($\times 10^{-8}$) &
$0.038$	&
$0.029$	&
$0.022$	&
$0.018$	&
$0.016$	&
$0.013$	&
$0.012$	&
$0.010$	&
$0.009$	\\ \hline
\end{tabular}
\caption{The calculated muon antineutrino flux per proton on target (POT) for the data included in this analysis}
\end{table*}
\endgroup

\fi

\end{document}